\documentclass[journal,onecolumn]{IEEEtran}
\usepackage{cite}
\usepackage{amsmath,amssymb,amsfonts}
\usepackage{algorithmic}
\usepackage{graphicx}
\usepackage{textcomp}

\usepackage{color}
\usepackage{circuitikz}
\usepackage{xcolor}
\usepackage{array}
\tikzset{>=latex} 
\definecolor{myblue}{rgb}{0, 0.4470, 0.7410}
\definecolor{mygray}{rgb}{0.7, 0.7, 0.7}
\usepackage{balance}
\usepackage{breqn}

\def\BibTeX{{\rm B\kern-.05em{\sc i\kern-.025em b}\kern-.08em
    T\kern-.1667em\lower.7ex\hbox{E}\kern-.125emX}}

\begin{document}
\title{Super Realized Gain Antenna Array}

\author{
Donal Patrick Lynch, \IEEEmembership{Student Member,~IEEE}
Manos M. Tentzeris, \IEEEmembership{Fellow, IEEE} 
Vincent Fusco, \IEEEmembership{Fellow, IEEE} 
\\and  Stylianos D. Assimonis \IEEEmembership{Member,~IEEE}
\thanks{Donal Patrick Lynch, Vincent Fusco and Stylianos D. Assimonis are with the Institute of
Electronics Communications and Information Technology, Queen’s University
Belfast, BT3 9DT Belfast, U.K. (e-mail: 
dlynch27@qub.ac.uk;
v.fusco@ecit.qub.ac.uk;
s.assimonis@qub.ac.uk).}
\thanks{Manos M. Tentzeris is with the Georgia Institute of Technology, School of Electrical and Computer Engineering, Atlanta, GA 30332 USA (e-mail:
etentze@ece.gatech.edu)}
}

\maketitle

\begin{abstract}

In this study, we investigate and fabricate a superdirective antenna array composed of strip dipole elements operating at a frequency of $3.5$ GHz. The spacing, dimensions, and phase difference of the elements are optimized to achieve a super realized gain antenna with a theoretical efficiency of $98.8\%$ and computed efficiency of $99.3\%$.
By employing an element spacing of $0.2\lambda$, the end-fire antenna array demonstrates a maximum theoretical realized gain of $6.4$ dBi, and a maximum computed realized gain of $6.3$ dBi.
Significantly, our proposed superdirective antenna array distinguishes itself from existing approaches by achieving high directivity, high radiation efficiency, and impedance matching to $50$ $\Omega$ solely through careful adjustments in the strip dimensions and the inter-element phase.
This eliminates the need for additional impedance matching networks, amplifiers, or attenuators.
%

\end{abstract}

\begin{IEEEkeywords}
Antennas, Antenna arrays, Microstrip antenna arrays, Dipole antennas, Directive antennas, Superdirective antenna arrays
\end{IEEEkeywords}

\section{Introduction}
\label{sec:introduction}
\IEEEPARstart{T}{he} advent of modern mobile communication networks, including fifth-generation (5G), sixth-generation (6G), and beyond, has been driven by the ever-increasing demand for faster download speeds and low latency, enabling seamless connectivity to work and social digital platforms. With the exponential growth of data-intensive applications and the need for reliable connectivity, these systems have become a necessary response to meet the evolving requirements of today's digital society. One of the critical aspects of these advanced communication technologies is the deployment of efficient and advanced antenna systems \cite{assimonis2021millimeter,Hu2023Additively} to support the enhanced capabilities of these networks.

In the context of 5G \cite{Eid2021}, which is currently the most prevalent mobile communication technology, the frequency band most widely used for applications falls within the range of $3.3$ to $4.2$ GHz. 
The selection of the sub-6 GHz range for 5G deployment is driven by the desire to strike a balance between coverage and capacity. However, a notable challenge faced by 5G antenna systems is their relatively large size, which gives rise to intricate and complex geometries. These larger and more complex antenna systems pose difficulties in the fabrication process, making their practical implementation more challenging. The increased complexity adds to the manufacturing complexity and costs, thereby necessitating innovative design and fabrication techniques to overcome these practical implementation challenges in deploying 5G antenna systems effectively.


To address this challenge, researchers and engineers have been exploring alternative solutions, and one promising approach is the use of \textit{superdirective} antenna arrays (SDAs). SDAs offer compact geometries and higher directivity compared to traditional uniform antenna arrays. This increased directivity is achieved through the close placement of antenna elements, which results in strong coupling between them. Several studies, including the work of Uskov \cite{uzkov} and others  
\cite{Yaru1951,Uzsoky1956,Harrington1958,Altshuler2005monopole,Morris2005Superdirectivity,Ivrlac2010High,Kim2012Superdirective,Marzetta2019Super,dovelos2022superdirective,Han2022Coupling,dovelos2022superdirective2,Tornese2022Gain,han2023superdirective,Ziolkowski2023Superdirective},
have demonstrated that a typical linear antenna array with $N$  elements yields a maximum directivity of $N^2 + 2N$ as the inter-element distance tends to zero. In general, this superdirectivity is higher than that of a corresponding uniform antenna array with the same number of elements. Furthermore, a superdirective antenna array is end-fire, meaning that the radiation pattern of the array is directed predominantly along the axis perpendicular to the array's elements.

By leveraging SDAs in 5G systems, it becomes possible to enhance the received power and extend the communication distance. This, in turn, leads to improved power efficiency and ensures reliable connectivity, even in challenging environments. The compact nature of SDAs also helps address the issue of large antenna size, enabling easier integration into various devices and infrastructure.

However, it is important to note that SDAs also face certain disadvantages and impediments to practical implementation. One significant drawback is the presence of high ohmic losses, which reduce radiation efficiency and antenna gain. These losses limit the overall performance of the antenna system and affect its ability to efficiently transmit and receive signals. Additionally, the impedance characteristics of SDA elements often exhibit high reactance, making it challenging to achieve impedance matching at the standard $50$ $\Omega$ \cite{Couraud2021Internet}, thereby complicating the practical implementation of superdirective arrays.



In this paper, our first objective is to conduct a theoretical study on an antenna array consisting of two dipoles, with a focus on superdirectivity. This study can be readily extended to an antenna array consisting of $N$ elements. In contrast to existing research, we prioritize the analysis of the realized gain rather than solely examining directivity or gain. We make this choice because the realized gain factor takes into account both ohmic and return losses. Additionally, we propose implementing dipoles with slightly different lengths and radii (or widths), excited by signals of equal magnitude but different phases. This innovative approach allows us to achieve impedance matching of the antenna array elements to $50$ $\Omega$, eliminating the need for additional impedance matching networks, hence, this technique reduces ohmic losses and enhances radiation efficiency. Moreover, by adopting this design concept, there is no need for active amplifiers or attenuators to regulate the magnitude of the excitation signal, resulting in a significant reduction in the antenna design process. Additionally, the power efficiency of the superdirective antenna system is enhanced by minimizing power consumption. As the next step, we conduct a comprehensive numerical analysis using full-electromagnetic simulation. Finally, we proceed to fabricate and measure the proposed antenna array with super realized gain. The measured results closely align with the simulated and theoretical findings.

\section{Antenna Design}
\subsection{Theoretical Analysis}
\begin{figure}[!t] 
\centering
\def\myLa{2}
\def\myWa{5}
\def\myLb{2.25}
\def\myWb{6}
\def\myD{3}
\def\myB{1.35}
\def\myG{0.05}
\def\myI{0.5}
\def\myLeg{0.35}
\def\myLegOff{1.1}
\def\myVs{1}
\def\myC{5}

\begin{circuitikz}
\coordinate (start) at (0,0);
\coordinate (P) at (\myD,\myB);
\coordinate (A) at (\myD,-\myB*0.8);
\coordinate (B) at (0,-\myB*0.8);
\draw[line width=0.5pt, rounded corners, color=black, ->]
(start) -- ++(1.6*\myD,0) node[right] {$\rho$};
\draw[line width=0.5pt, rounded corners, color=black, ->]
(start) -- ++(0,1.5*\myLa) node[above] {$z$};
\draw[line width=\myWa pt, rounded corners, color=mygray]
(start)++(0,\myG) -- (0,\myLa)
(start)++(0,-\myG) -- (0,-\myLa);
\draw[line width=\myWb pt, rounded corners, color=mygray]
(start)++(\myD,\myG) -- (\myD,\myLb)
(start)++(\myD,-\myG) -- (\myD,-\myLb)
;
\draw[line width=0.35pt, rounded corners, color=black, - ]
(start) -- (P) node[midway, above] {$R_c^{(21)}$}
node[right] {$E(z)$}
(start)++(0,-\myLa) -- (P) node[midway, left] {$R_b^{(21)}$}
(start)++(0,\myLa) -- (P) node[midway, above] {$R_a^{(21)}$};
\draw[line width=0.5pt, rounded corners, color=black, ->]
(A)++(0,-\myI/2) -- ++(0,\myI) node[midway, left] {$I_2(z)$};
\draw[line width=0.5pt, rounded corners, color=black, ->]
(B)++(0,-\myI/2) -- ++(0,\myI) node[midway, left] {$I_1(z)$};
\draw[line width=0.5pt, rounded corners, color=black, Bar-Bar ]
(start)++(-\myLegOff,-\myLa) -- ++(0,2*\myLa) node[midway, left] {$L_1$};
\draw[line width=0.5pt, rounded corners, color=black, Bar-Bar]
(start)++(\myD+\myLegOff,-\myLb) -- ++(0,2*\myLb) node[midway, above right] {$L_2$};
\draw[line width=0.5pt, rounded corners, color=black, Bar-Bar]
(start)++(-\myWa/2 pt,-\myLa-\myLegOff/2) -- ++(\myWa pt,0) node[right] {$2 a_1$};
\draw[line width=0.5pt, rounded corners, color=black, Bar-Bar]
(start)++(\myD cm -\myWb/2 pt,-\myLa-\myLegOff/2/0.9) -- ++(\myWb pt,0) node[right] {$2 a_2$};
\draw[line width=0.5pt, rounded corners, color=black, Bar-Bar]
(start)++(0,-\myLb-\myLegOff*0.9) -- ++(\myD,0) node[midway, above] {$d$};
\end{circuitikz}
\caption{An array of wire dipole antennas with an inter-element distance of $d$, lengths of elements $L_1$ and $L_2$, and radii of elements $a_1$ and $a_2$.}
\label{Fig:Geometry0}
\end{figure}
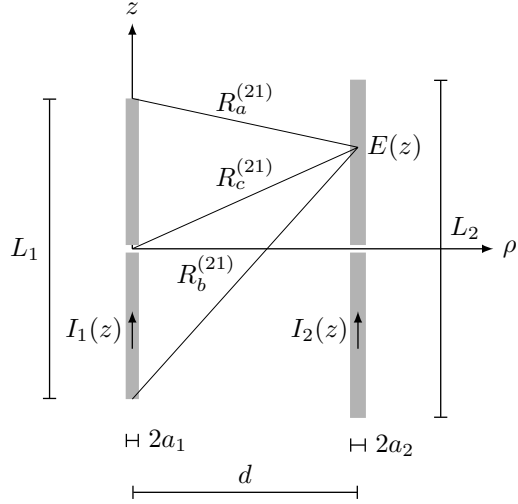

When antennas are located in close proximity, the effect of mutual coupling between them cannot be ignored. The mutual impedance serves as an indicator of the degree to which antenna cross interaction proximity effects are present. This section presents the theoretical analysis of a two-element antenna array, as depicted in Fig. \ref{Fig:Geometry0}. The array consists of linear wire dipoles arranged in a parallel, side by side format along the $\rho$-axis at a distance $d$ from each other, with centers at positions $(x_i, y_i)$, lengths $L_i$, radii $a_i$, input voltages and currents $V_i$ and $I_i$, respectively, where $i=1,2$. In order to evaluate the mutual and self impedances $Z_{ij}$ of the antenna array, where $i, j = 1, 2$, we consider it as a two-port network. Therefore, in general, when both antennas are excited, the relationship between the driving voltages and input currents are expressed as follows \cite{pozar2011microwave}:
\begin{equation}\label{eqZmat}
\begin{bmatrix}
V_1\\
V_2
\end{bmatrix}
=
\begin{bmatrix}
Z_{11} & Z_{12}\\
Z_{21} & Z_{22}
\end{bmatrix}
\begin{bmatrix}
I_1\\
I_2
\end{bmatrix}
\Leftrightarrow
\mathbf{v} = \mathbf{Z_n} \, \mathbf{i_n}
\end{equation}
The impact of the first dipole on the second dipole is represented by the mutual impedance $Z_{21}$, which is defined as \cite{Orfanidis2016}:
\begin{equation}\label{Z21}
Z_{21} = 
\frac{V_2^{oc}}{I_1} 
\end{equation}
Thus, the mutual impedance $Z_{21}$ is defined as the ratio of the induced open-circuited voltage at the terminals of the second dipole when only the first dipole is driven, and vice versa for $Z_{12}$. Please note that according to reciprocity, $Z_{21} = Z_{12}$. The induced open-circuited voltage is given by \cite{Orfanidis2016}:
\begin{equation}\label{V21oc}
V_2^{oc}
= -\dfrac{1}{I_2} \int_{-l_2}^{l_2}
E_z \left( z \right) I_2 \left( z \right) \mathrm{d}z,
\end{equation}
where $l_2 = L_2/2$ and $E_z(z)$ is the electric field caused by the driven first dipole on the second dipole. To calculate $E_z(z)$, we need to define the currents that flow through the dipoles. We assume that the dipoles have lengths close to half wavelength, so we can use a sinusoidal current distribution in this analysis. Thus, we have:
\begin{align}\label{I2}
I_2\left( z \right) = I_2 
\frac{\sin
\left[  
k \left( l_2-\left| z\right| \right)  
\right]  
}{\sin
\left[ 
k \, l_2
\right] 
},
\quad
\left| z\right| \le l_2,
\end{align}
where $I_2\in \mathbb{C}$ is the input current of the second dipole, $k = 2\pi/\lambda$ is the wavenumber, and $\lambda$ is the wavelength. Please note that $I_1(z)$ can also be given by \eqref{I2} by setting $I_2\rightarrow I_1$ and $l_2 \rightarrow l_1 = L_1/2$. The electric field along the second antenna is given by \cite{Orfanidis2016}:
\begin{dmath}\label{Ez}
E_z \left( z \right)  = -j  \dfrac{\eta_0  I_1}{4\pi  \sin \left[k \, l_1\right]   }
\left(    
\frac{e^{\,- j k R_a^{(21)}}}{R_a^{(21)}} +
\frac{e^{\,- j k R_b^{(21)}}}{R_b^{(21)}} -
2\cos \left[  k \, l_1 \right] 
\frac{e^{\,- j k R_c^{(21)}}}{R_c^{(21)}}
\right),
\end{dmath}
where $\eta_0$ is the characteristic impedance of free space and $z \in [-l_2,l_2]$, 
and
\begin{align}\label{RaRbRc}
	R^{(21)}_a &= \sqrt{d^2 + \left( z - l_1 \right) ^2}\nonumber\\
	R^{(21)}_b &= \sqrt{d^2 + \left( z + l_1 \right) ^2}\\
	R^{(21)}_c &= \sqrt{d^2 + z^2} \nonumber
\end{align}
By substituting  \eqref{V21oc}-\eqref{RaRbRc} into  \eqref{Z21}:
\begin{equation}\label{Z21b}
	Z_{21} = j \dfrac{\eta_0  }{4\pi  \sin \left[k \, l_1\right] \sin \left[k \, l_2\right]  }
	\int_{-l_2}^{l_2}
	A_{21}\left( z \right)
	\mathrm{d}z
\end{equation}
where,
\begin{dmath}
	A_{21}\left( z \right)\label{A}
	=
	\left(    
	\frac{e^{\,- j k R^{(21)}_a}}{R^{(21)}_a} +
	\frac{e^{\,- j k R^{(21)}_b}}{R^{(21)}_b} -
	2\cos \left[  k \, l_1 \right] 
	\frac{e^{\,- j k R^{(21)}_c}}{R^{(21)}_c}
	\right)
	\sin
	\left[  
	k \left( l_2-\left| z\right| \right)  
	\right].
\end{dmath}
Note that the integral in  \eqref{Z21b} does not have an analytical solution. Therefore, we utilized numerical integration techniques, specifically global adaptive quadrature, to accurately evaluate the integral \cite{shampine2008vectorized}.

To obtain the near-field on the surface of the first dipole, we set $d \rightarrow a_{1}$ and $l_2 \rightarrow l_1$ in  \eqref{RaRbRc} because the integral is now estimated on the first dipole, and not on the second. The resulting expression is:
\begin{equation}\label{Z11}
	Z_{11} = j \dfrac{\eta_0  }{4\pi  \sin^2 \left[k \, l_1\right]  }
	\int_{-l_1}^{l_1}
	A_{11}\left( z \right)
	\mathrm{d}z,
\end{equation}
where now,
\begin{dmath}
	A_{11}\left( z \right)\label{A2}
	=
	\left(    
	\frac{e^{\,- j k R^{(11)}_a}}{R^{(11)}_a} +
	\frac{e^{\,- j k R^{(11)}_b}}{R^{(11)}_b} -
	2\cos \left[  k \, l_1 \right] 
	\frac{e^{\,- j k R^{(11)}_c}}{R^{(11)}_c}
	\right)
	\sin
	\left[  
	k \left( l_1-\left| z\right| \right)  
	\right]
\end{dmath}
and
\begin{align}\label{RaRbRc2}
	R^{(11)}_a &= \sqrt{a_1^2 + \left( z - l_1 \right) ^2}\nonumber\\
	R^{(11)}_b &= \sqrt{a_1^2 + \left( z + l_1 \right) ^2}\\
	R^{(11)}_c &= \sqrt{a_1^2 + z^2}. \nonumber
\end{align}
Similar analysis can be applied to estimate $Z_{22}$. 
%
	
With the given driven voltages $\mathbf{v}$, solving equation \eqref{eqZmat} provides the input currents $\mathbf{i_n}$, which are used to define the sinusoidal currents $I_i(z)$ based on  \eqref{I2}. Thus,
\begin{equation}\label{In}
	\mathbf{i_n} = \mathbf{Z_n}^{-1} \, \mathbf{v}.
\end{equation}
where, $\left(~\right)^{-1} $ denotes the inverse matrix.

By determining the currents $\mathbf{i_n}$, the radiation pattern of the array can be obtained, and the radiation intensity can be expressed as:
	\begin{dmath}
		U\left( \mathbf{i_n},\theta, \phi \right) = 
		\frac{\eta_0}{8 \, \pi^2} 
		\left| 
		\sum_{i=1}^{N} 
		I_i
		\frac{\cos\left[ k \, l_i \cos \theta \right] - \cos \left[ k \, l_i \right] }{\sin \left[ k \, l_i\right] \sin \theta}
		e^{
		\, j \vec{k}\cdot\vec{d_i} 
		}
		\right|^2
	\end{dmath}
	where $N$ is the number of antenna array elements (in this example $N=2$ since     we have two dipoles), $\vec{k} = k \hat{r}$ is the wavevector, where
	\begin{equation}
		\hat{r} = \sin{\theta} \cos{\phi} \, \hat{x} + \sin{\theta} \sin{\phi} \, \hat{y} + \cos{\theta} \, \hat{z},
	\end{equation}
	is the unit vector in spherical coordinates, and $\vec{d_i} = x_i , \hat{x} + y_i , \hat{y} + z_i , \hat{z}$ is the vector that indicates the position of the dipoles. Hence, the directivity is given by
	\begin{equation}\label{D01}
		D \triangleq 4\pi \frac{U\left( \mathbf{i_n},\theta, \phi \right)}{P_{r}},
	\end{equation}
	where
	 \begin{equation}\label{Pr01}
	 	P_{r} \triangleq \int_{\phi=0}^{2\pi} \int_{\theta=0}^{\pi} U \left( \mathbf{i_n},\theta, \phi \right) \sin \theta \, \mathrm{d}\theta \mathrm{d}\phi,
	 \end{equation}
	 is the total radiated power. The radiated power represents a portion of the input power to the two-port system and is defined as:
	 \begin{equation}
	 	P_{in} \triangleq P_r + P_l,
	 \end{equation}
  where $P_l$ represents the ohmic losses on the dipoles. In the calculation of directivity, it is assumed that there are no ohmic losses on the antenna array, and therefore all the input power is radiated. In this scenario (i.e., $P_l\rightarrow 0$), it can be presumed that $P_{\text{in}} = P_{\text{r}}$, as stated in \cite{dovelos2022superdirective}:
 	\begin{align}\label{Pr02}
 		P_{r} &= 
 		\frac{1}{2}
 		\mathrm{Re}
 		\left\lbrace
 		\mathbf{i_n}^{H} \mathbf{Z_{n}} \, \mathbf{i_n}
 		\right\rbrace\nonumber\\
 		&=
 		\frac{1}{2}
 		\mathrm{Re}
 		\left\lbrace
 		\left(\mathbf{Z_n}^{-1} \, \mathbf{v}\right) ^{H} \, \mathbf{v}
 		\right\rbrace{},
 	\end{align}
 where, $\left(~\right)^{H} $ denotes the Hermitian  transpose.

On the other hand, when  calculating the gain of an antenna array, it is essential to consider the ohmic losses associated with the wire dipoles.
In general, the gain of an antenna is given by:
\begin{equation}\label{G01}
G \triangleq 4\pi \frac{U\left( \mathbf{i_l},\theta, \phi \right)}{P_{in}} 
	= 4\pi \frac{U\left( \mathbf{i_l},\theta, \phi \right)}{P_{r}+P_{l}},
\end{equation}
where now $P_l \ne 0$. Ohmic losses are a result of the \textit{skin effect} \cite{balanis2015antenna}. Based on this effect, we can derive the loss resistance per unit length on the $i$-th conductive wire dipole as:
\begin{equation}
r_{l,i} = \frac{1}{2 a_i} \sqrt{\frac{f \mu_0}{\pi \sigma}},
\end{equation}
where $f$, $\mu_0 = 4\pi \times 10^{-7}$ H/m, and $\sigma$ are the operating frequency, magnetic permeability of free space, and wire conductivity, respectively. Thus, for the current distribution of \eqref{I2},
\begin{equation}\label{loss}
R_{l,i} = r_{l,i} \int_{-l_i}^{l_i} \left| \frac{I_i\left(z\right)}{I_i}\right|^2 \, \mathrm{d}z = 
\frac{k L_i - \sin \left[k L_i\right] }{4 k a_i \sin^2 \left[k \frac{L_i}{2}\right] }
\sqrt{\frac{f \mu_0}{\pi \sigma}}.
\end{equation}
Additionally, the relationship between the driving voltages and the input currents is now expressed as follows:
\begin{equation}\label{Zlmat}
\mathbf{v} = \mathbf{Z_l} \, \mathbf{i_l} = \left( \mathbf{Z_{n}}+\mathbf{R_{l}} \right) \, \mathbf{i_l},
\end{equation}
where $\mathbf{R_{l}} = \mathrm{diag}(R_{l,1},\ldots,R_{l,N})$, and $\mathbf{i_l}$ is the matrix of the input currents of the lossy network. Hence, similarly to \eqref{In}:
\begin{equation}\label{Il}
\mathbf{i_l} = \left( \mathbf{Z_n}+ \mathbf{R_{l}} \right) ^{-1} \mathbf{v}.
\end{equation}
 	
The input power is now given by:
\begin{align}\label{Pr03}
P_{in} &= 
\frac{1}{2}
\mathrm{Re}
\left\lbrace
\mathbf{i_l}^{H} \left(\mathbf{Z_{n} + \mathbf{R_{l}}}\right)  \mathbf{i_l} 
\right\rbrace
\nonumber\\
&=
\frac{1}{2}
\mathrm{Re}
\left\lbrace
\left(\left( \mathbf{Z_n}+ \mathbf{R_{l}} \right) ^{-1} \mathbf{v}\right) ^{H} \mathbf{v}
\right\rbrace.
\end{align}
The definition of the directivity and gain of an antenna array incorporates the corresponding power density. The power density is determined by the input currents, which do not consider the ohmic losses on the radiating elements when estimating directivity, but do take into account the ohmic losses when estimating gain. Consequently, the power density for directivity (i.e., $U(\mathbf{i_n},\theta_o, \phi_o)$) differs from the power density for gain (i.e., $U(\mathbf{i_l},\theta_o, \phi_o)$). Additionally, the radiation efficiency, defined as the maximum gain divided by the maximum directivity, is represented by the following equation:
\begin{align}\label{etanew}
	\eta &\triangleq \frac{G}{D}
	=
	\frac{U_{\mathrm{max}}\left( \mathbf{i_l},\theta_o, \phi_o \right) \, P_{r} }{ U_{\mathrm{max}}\left( \mathbf{i_n},\theta_o, \phi_o \right) \, P_{in} },
\end{align}
which incorporates the maximum radiation density for the lossless case (i.e., $U_{\mathrm{max}}(\mathbf{i_n},\theta_o, \phi_o)$) and the lossy case (i.e., $U_{\mathrm{max}}(\mathbf{i_l},\theta_o, \phi_o)$).

It is noted that in the literature, the radiation efficiency is often defined as the ratio of the radiated power to the input power, expressed as:
\begin{equation}\label{etanew2}
    \eta=\frac{P_{r}}{P_{in}}.
\end{equation}
Equation \eqref{etanew2} follows from equation \eqref{etanew}, assuming an identical power density in both directivity and gain estimation. However, it is crucial to note that power density depends on the input currents, which vary when estimating directivity and gain. The input currents account for ohmic losses in the gain estimation, but this consideration is absent in the directivity estimation. Consequently, this disparity leads to inaccuracies when using \eqref{etanew2} to calculate radiation efficiency.
%
%
For instance, when considering the scenario where copper wires with lengths $L_1=L_2=\lambda/2$, radii $a_1=a_2=\lambda/1001$, are placed side-by-side at a distance $d=\lambda/2$, and driven by voltages $V_1=V_2=1$ V, the calculated value of $\eta$ based on \eqref{etanew} is $99\%$. 
On the other hand, applying \eqref{etanew2} yields a different value of $100.94\%$, clearly demonstrating the inaccuracies caused by the application of this simplified formula.


Another important parameter is the realized gain, which is defined as the product of the port efficiency $\eta_{\mathrm{port}}$ and the gain, and thus:
\begin{equation}\label{RG01}
	G_R \triangleq \eta_{port} \, G,
\end{equation}
where, 
\begin{equation}\label{etaport}
	\eta_{port} = 
 1 -  \left|\Gamma_a \right| ^2 = 
 \frac{ \mathbf{v}^{H} \left(\mathbf{I}-\mathbf{S}^{H}\mathbf{S}\right) \mathbf{v} }{\mathbf{v}^{H}\mathbf{v}},
\end{equation}
where $\Gamma_a$ is the total active reflection coefficient \cite{Manteghi2003Broadband,Moradi2022Some}, $\mathbf{S}$ is the $S$-parameter matrix of the two-port network, calculated at the reference impedance of $Z_0=50$ $\Omega$, and $\mathbf{I}$ is an identity matrix with the same dimension as $\mathbf{S}$.

The antenna array was optimized to achieve superdirectivity. The goal was to maximize the directivity, gain (which considers ohmic losses, equivalently radiation efficiency), and realized gain (which accounts for both radiation efficiency and return losses at $50$ $\Omega$ in our case) by varying the inter-element distance from $0.1\lambda$ to $0.5\lambda$, where $\lambda$ is the operating frequency wavelength (assumed to be $3.5$ GHz for sub-6 GHz 5G systems). The analysis is based on theoretical calculations and formulas, i.e., on  \eqref{D01}, \eqref{G01}, and \eqref{RG01}. Design parameters include the lengths ($L_1, L_2$), radii ($a_1, a_2$), and inter-element phase difference ($\Delta\phi$). In contrast to the state-of-the-art approach \cite{Altshuler2005monopole,dovelos2022superdirective,dovelos2022superdirective2,Han2022Coupling}, we fixed the driven voltages' magnitude at $1$ V/m to avoid using additional components like amplifiers or attenuators.

\begin{figure}[!t]
\begin{minipage}{.48\linewidth}
  \centering
  \includegraphics[height=.63\linewidth]{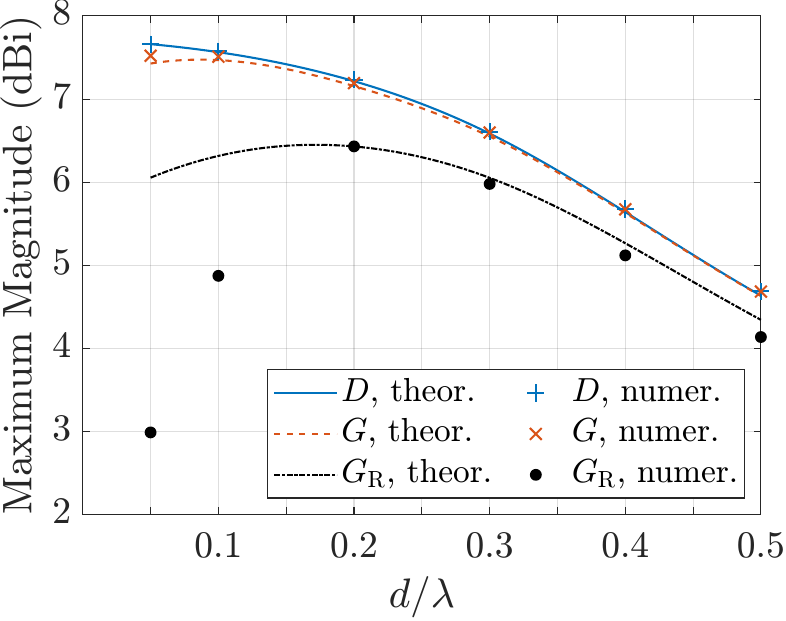}
  \caption{The end-fire directivity, gain, and realized gain of the optimal antenna array in terms of dB-scaling are analyzed in relation to the inter-element distance (analytical and numerical results).}
  \label{Fig:DirGainRGainVSd}
\end{minipage} 
\hspace{5mm}
\begin{minipage}{.48\linewidth}
  \centering
  \includegraphics[height=.63\linewidth]{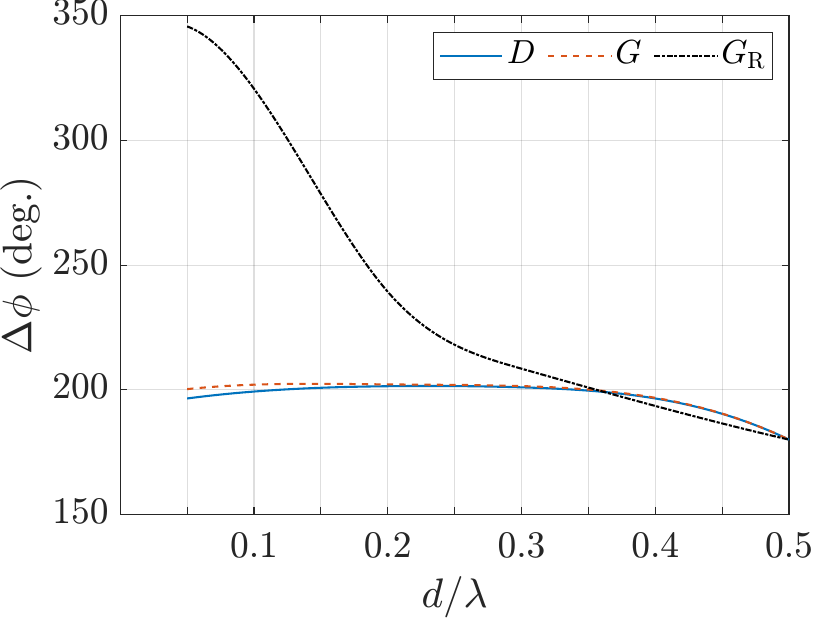}
  \caption{Inter-element phase difference for maximum end-fire directivity, gain, and realized gain of a two-wire dipole antenna array versus inter-element distance. The remaining design parameters (i.e., $L_1$, $L_2$, $a_1$, and $a_2$) are held constant at their optimal values.}
  \label{Fig:DeltaPhi_dir_gain_Rgain_001}
\end{minipage}
\end{figure}

Particle Swarm Optimization (PSO) was employed as the optimization method for the solution of the problem:
%
\begin{equation}
\begin{aligned}
\underset{
\left\lbrace L_1,L_2,a_1,a_2,\Delta\phi \right\rbrace 
}{\text{Maximize}}
& \quad f \left(L_1,L_2,a_1,a_2,\Delta\phi\right) \\
\text{subj. to:} & \quad L_1, L_2 \in \left[0.4\lambda, 0.6\lambda\right],\\
         & \quad a_1, a_2 \in \left[\lambda/2001, \lambda/201\right],\\
         & \quad \Delta\phi \in \left[0^{\circ}, 360^{\circ}\right],
\end{aligned}
\end{equation}
where, function $f$ represents either the directivity \eqref{D01}, the gain \eqref{G01} or the realised gain \eqref{RG01}.
Each inter-element distance had an optimal set of design parameters for maximum directivity, gain, and realized gain. 
The results in Fig. \ref{Fig:DirGainRGainVSd} reveal a significant increase in directivity as the inter-element distance approaches zero, indicating superdirectivity. The gain reaches its maximum when the inter-element distance is approximately $0.1\lambda$, accounting for ohmic losses. Similarly, the realized gain, which considers return losses at $50\,\Omega$, peaks at an inter-element distance of around $0.2\lambda$ (specifically at $0.17\lambda$). In contrast, the directivity trend alone suggests enhancement as $d$ tends to zero with appropriate excitation signals, this finding does not account for ohmic losses or return losses. Moreover, the directivity calculated at $d=0.2\lambda$ is $7.3\,\text{dBi}$. Additionally, the directivity estimated in \cite{Altshuler2005monopole} for two isotropic elements is $3.5$ when using linear scaling. In our specific scenario, considering dipoles of length close to half-wavelength and a theoretical maximum directivity of $1.67$, the predicted directivity is $10\log_{10}(1.67 \cdot 3.5) \approx 7.7$ dBi, which closely aligns with our  findings.

Fig. \ref{Fig:DeltaPhi_dir_gain_Rgain_001} depicts the optimal inter-element phase difference ($\Delta\phi$) for achieving maximum directivity, gain, and realized gain as a function of inter-element distance. While $\Delta\phi$ displays significant variation for realized gain, it remains approximately $200^\circ$ for directivity and gain, even up to an inter-element distance of $0.4\lambda$. This observation suggests that the directivity and gain are relatively insensitive to the phase setting.

\begin{table*}[h] 
\centering
\renewcommand{\arraystretch}{1.3} 
\caption{Optimum Design Parameters: Theoretical Analysis at 3.5 GHz} 
\label{Table01}
\begin{tabular}{
c
|c
|c
|c
|c
|c
|c
|c
|c
}
  \hline
  $d/\lambda$ 
  & $L_1/\lambda$ 
  & $L_2/\lambda$ 
  & $a_1/\lambda$
  & $a_2/\lambda$ 
  & $\Delta\phi$ (deg.) 
  & $\eta_{port}$
  & $\eta~(\%)$
  & $G_R$ (dBi)
  
  \\
  \hline
  \hline
  $0.05$
  &$0.480$
  &$0.482$
  &$0.0050$
  &$0.005$
  &$345.8^{\circ}$
  &$0.763$
  &$94.2$
  &$6.1$
  \\
  \hline
  $0.1$
  &$0.473$
  &$0.467$
  &$0.0050$
  &$0.005$
  &$320.8^{\circ}$
  &$0.807$
  &$98.4$
  &$6.3$
  \\
  \hline
  $0.2$
  &$0.479$          
  &$0.452$          
  &$0.0015$          
  &$0.002$          
  &$239.3^{\circ}$
  &$0.925$
  &$98.8$
  &$6.4$
  \\
  \hline
  $0.3$    
  &$0.474$          
  &$0.437$         
  &$0.0009$          
  &$0.005$          
  &$208.5^{\circ}$
  &$0.978$
  &$99.2$
  &$6.1$
  \\
  \hline
  $0.4$    
  &$0.466$          
  &$0.440$          
  &$0.0012$          
  &$0.005$          
  &$193.5^{\circ}$
  &$0.997$
  &$99.5$
  &$5.3$
  \\
  \hline
  $0.5$    
  &$0.448$          
  &$0.448$          
  &$0.0050$
  &$0.005$
  &$180^{\circ}$
  &$0.987$        
  &$99.8$
  &$4.3$  
  \\
  \hline
  \hline
\end{tabular}
\end{table*}

Finally, Table \ref{Table01} provides insight into the lengths ($L_1, L_2$) and radii ($a_1, a_2$) yielding the maximum realized gain as a function of the inter-element distance. Notably, the lengths of the wire-dipoles consistently remain below half-wavelength for inter-element distances up to $0.5\lambda$. This suggests that optimizing the lengths within this range is crucial for achieving high realized gain. Additionally, the radii of the wire-dipoles appear to approach the upper limit of $\lambda/201 \approx 0.005\lambda$ in most cases. This observation implies that increasing the radius of the wires may lead to even higher realized gain. However, to maintain the assumption of a valid sinusoidal current distribution over the dipoles, we adhered to the empirical rule of keeping the wires as thin as possible.

According to Harrington's study \cite{Harrington1958}, the maximum directivity, $D_{\text{max}}$, of a lossless antenna that completely fills a sphere with radius $R$ is given by the equation:
\begin{equation}
D_{\text{max}} = (kR)^2 + 2kR
\end{equation}
In our specific case, with an inter-element distance of $0.2\lambda$, the resulting radius $R$ is approximately $22$ mm. Applying Harrington's findings, the maximum directivity of such an antenna is $7.7$ dBi, which aligns with our expectations. Additionally, the achieved realized gain reaches a maximum of $6.4$ dBi, indicating close proximity to this theoretical upper limit.

These analytical findings suggest that it is feasible to implement a practical super-realized gain antenna by carefully designing the dipoles in the array.

\subsection{Numerical Analysis}

Concluding the theoretical analysis presented in the previous Section, we validated our findings through numerical analysis. Specifically, we first applied the method of moments using the Antenna Toolbox of MATLAB \cite{matlab2022}. We modeled the dipoles using the \textit{dipoleCylindrical} function and the array using the \textit{linearArray} function. The conductor was defined using the \textit{metal} function, with a conductivity of $5.8\times 10^7$ S/m and a thickness of $35$ $\mu$m. We used the same optimum design parameters as in the theoretical study. The results are depicted in Fig. \ref{Fig:DirGainRGainVSd}.
Regarding the directivity, there is perfect agreement between the theoretical and numerical results from $0.05\lambda$ to $0.5\lambda$. A good agreement is also observed for the gain case. However, as the dipoles get closer to each other or become thicker, the agreement for the realized gain decreases. This has an impact on the estimation of \eqref{loss}, which in turn affects the resulting impedance matrix $\mathbf{Z_l}$ in \eqref{Zlmat}. The impedance matrix is used to calculate the $S$-parameters in \eqref{etaport}. 
Please note that the numerical method estimates the current distribution on the surface of a cylinder with radius $a_i$, which differs from the linear distribution predicted in the theoretical analysis based on  \eqref{I2}. Therefore, as the radius increases, the assumption of \eqref{I2} becomes less accurate.

One of the main objectives of this work is to construct a superdirective antenna array. Therefore, although wire dipoles offer analytical expressions and their theoretical study is feasible, we have chosen to focus on strip dipoles instead due to the advantages offered by the fabrication process. For example, fabricating wire dipoles with accurate inter-element distance is challenging compared to the fabrication of strip dipoles with accurate widths. Strip dipoles can be easily manufactured by etching metal traces on PCB substrates, resulting in structurally robust configurations, and accurate geometries. Additionally, strip dipoles often exhibit a wider bandwidth compared to wire dipoles \cite{balanis2015antenna}.

\begin{figure}[!t] 
\centering
\def\myLa{4cm}
\def\myWa{0.3cm}
\def\myLb{5cm}
\def\myWb{0.5cm}
\def\myD{3cm}
\def\myG{0.8cm}
	
\def\myLeg{0.35cm}
\def\myLegOff{0.4cm}
	
\def\myVs{1cm}

\def\myC{5}

\begin{circuitikz}[scale = 0.875, american voltages]
\ctikzset{sources/scale = 0.65}
\draw[->] (-\myC,0,0) -- (0.8-\myC,0,0) node[right] {$y$};
\draw[->] (-\myC,0,0) -- (-0.6-\myC,-0.4,0) node[below] {$x$};
\draw[->] (-\myC,0,0) -- (0-\myC,1,0) node[above] {$z$};
\draw[fill=myblue] (-\myD/2-\myWa/2, -\myLa/2) rectangle (-\myD/2+\myWa/2, -\myG/2);
\draw[fill=myblue] (-\myD/2-\myWa/2,    \myG/2) rectangle (-\myD/2+\myWa/2, \myLa/2);
\draw[fill=myblue] ( \myD/2-\myWb/2, -\myLb/2) rectangle ( \myD/2+\myWb/2, -\myG/2);
\draw[fill=myblue] ( \myD/2-\myWb/2,    \myG/2) rectangle ( \myD/2+\myWb/2, \myLb/2);
\draw
(-\myD/2, \myG/2) to[V, l=$V_1$] (-\myD/2,-\myG/2)
( \myD/2, \myG/2) to[V, l=$V_2$] ( \myD/2,-\myG/2)
;
\draw[<-] (-\myD/2-\myWa/2-\myLegOff, -\myLa/2) -- (-\myD/2-\myWa/2-\myLegOff, -\myLeg);
\draw[->] (-\myD/2-\myWa/2-\myLegOff,  \myLeg) -- (-\myD/2-\myWa/2-\myLegOff,   \myLa/2);
\node at (-\myD/2-\myWa/2-\myLegOff, 0) {$L_1$};
\draw[<-] ( \myD/2-\myWb/2-\myLegOff, -\myLb/2) -- ( \myD/2-\myWb/2-\myLegOff, -\myLeg);
\draw[->] ( \myD/2-\myWb/2-\myLegOff,  \myLeg) -- ( \myD/2-\myWb/2-\myLegOff,   \myLb/2);
\node at ( \myD/2-\myWb/2-\myLegOff, 0) {$L_2$};
\draw[->] (-\myD/2-\myWa/2-\myLegOff, -\myLa/2-\myLegOff) -- (-\myD/2-\myWa/2, -\myLa/2-\myLegOff);
\draw[<-] (-\myD/2+\myWa/2, -\myLa/2-\myLegOff) -- (-\myD/2+\myWa/2+\myLegOff, -\myLa/2-\myLegOff);
\node at (-\myD/2-\myWa/2-2*1.2*\myLeg, -\myLa/2-\myLegOff) {$w_1$};
\draw[->] ( \myD/2-\myWb/2-\myLegOff, -\myLb/2-\myLegOff) -- ( \myD/2-\myWb/2, -\myLb/2-\myLegOff);
\draw[<-] ( \myD/2+\myWb/2, -\myLb/2-\myLegOff) -- ( \myD/2+\myWb/2+\myLegOff, -\myLb/2-\myLegOff);
\node at ( \myD/2-\myWb/2-2*1.2*\myLeg, -\myLb/2-\myLegOff) {$w_2$};
\draw[<-] (-\myD/2, \myLb/2+\myLegOff) -- (-\myLeg, \myLb/2+\myLegOff);
\draw[->] (\myLeg, \myLb/2+\myLegOff) -- ( \myD/2, \myLb/2+\myLegOff);
\node at (0, \myLb/2+\myLegOff) {$d$};
\end{circuitikz}
\caption{Strip dipole layout with inter-element distance $d$, lengths of elements $L_1$, $L_2$, widths of elements $w_1$, $w_2$, excitation signal applied to elements $V_1$, $V_2$.}
\label{Fig:Geometry}
\end{figure}
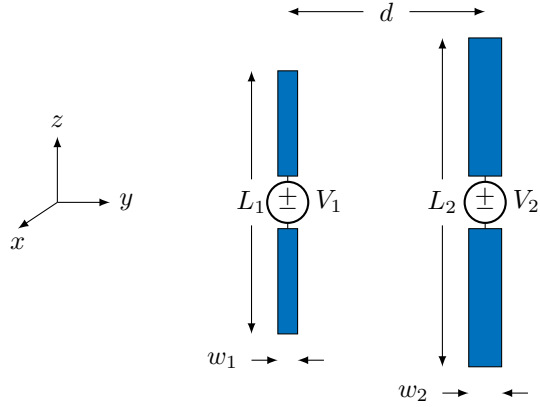

For comprehensive electromagnetic numerical simulations, we utilized the commercial solver CST Studio Suite 2022 \cite{CST2022}, specifically the time domain solver. This software allowed accurate modeling of the antenna array, analysis of its radiation pattern characteristics, and examination of the array's impedance. The simulated array is depicted in Fig. \ref{Fig:Geometry}. Copper material with a conductivity of $5.96 \times 10^7$ S/m and a thickness of $35$ $\mu$m was used for the metallic components. The strip dipoles were defined by their respective lengths, $L_1$ and $L_2$, widths, $w_1$ and $w_2$, and positioned at a distance of $d$ from each other. The dipoles were driven by input voltages $V_1$ and $V_2$. Discrete ports with $50$ $\Omega$ were used to model the excitation of the array. 
For practical reasons, the strips were modelled on a substrate based on RO4003C, with a thickness of $0.813$ mm, a dielectric constant of $\epsilon_r=3.55$, and a dissipation factor of $\tan\delta=0.0027$. The modeled substrate was assumed to have dimensions of $50 \times 50$ mm.

The antenna array was optimized to achieve superdirectivity by maximizing the realized gain while varying the inter-element distance from $0.05\lambda$ to $0.5\lambda$ (at $3.5$ GHz). The design parameters included the lengths $L_1$, $L_2$, the widths $w_1$, $w_2$, and the inter-element phase difference $\Delta\phi \in \left[0, 360^{\circ}\right]$ of the elements. To keep implementation complexity low, the magnitude of the driven voltages was again fixed at $1$ V/m.

\begin{figure}[!t] 
\begin{minipage}{.48\linewidth}
  \centering
  \includegraphics[height=.63\linewidth]{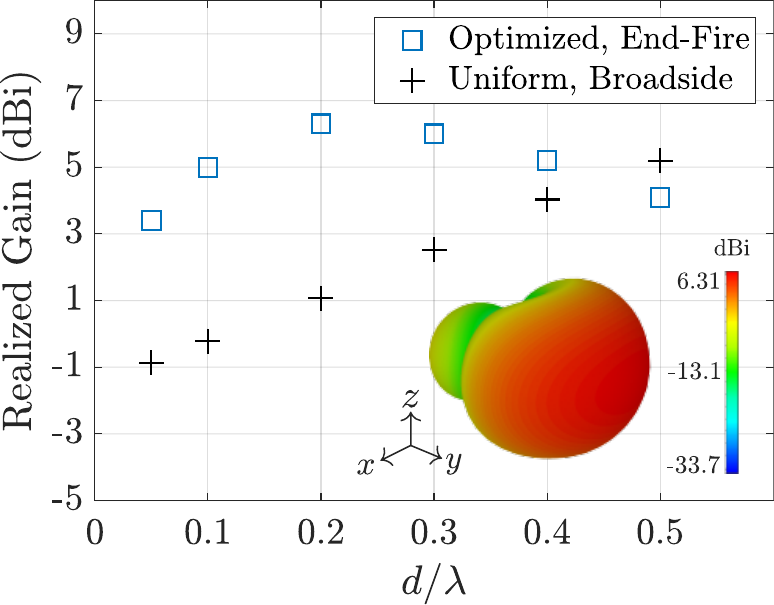}
  \caption{The antenna array's realized gain for the optimized and uniform cases versus inter-element distance at $3.5$ GHz (simulated results). The 3D radiation pattern for the optimized result when $d=0.2\lambda$ is also shown.}
  \label{Fig:RGainVSd}
\end{minipage} 
\hspace{5mm}
\begin{minipage}{.48\linewidth}
  \centering
  \includegraphics[height=.63\linewidth]{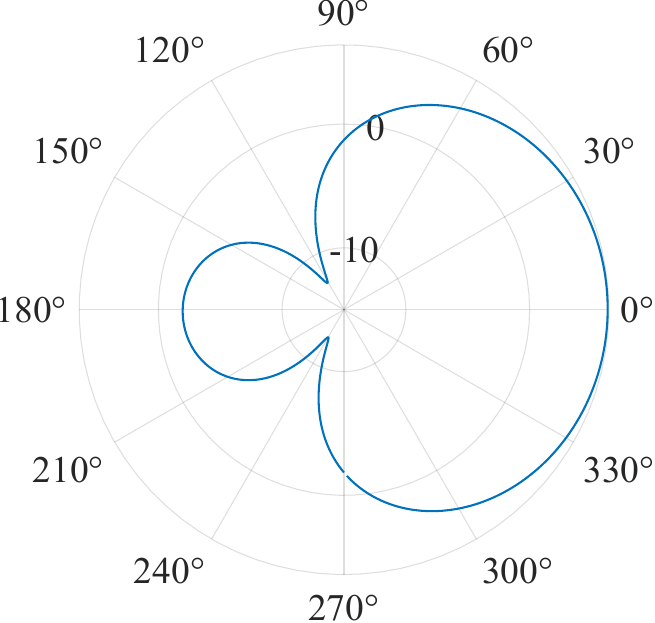}
  \caption{Simulated realized gain radiation pattern, horizontal plane ($H$-plane). Maximum occurs at $\phi = 0^{\circ}$ (i.e., end-fire antenna array). The front-to-back ratio at $3.5$ GHz is $8.44$~dB, and the angular width ($3$~dB) is $127^{\circ}$.}
  \label{Fig:RGainVSphi}
\end{minipage}
\end{figure}

PSO method was employed once again.
The outcomes are shown in Fig. \ref{Fig:RGainVSd}, which displays the maximum achieved antenna array realized gain as a function of the inter-element distance $d$, maximized to $6.3$ dBi at $d=0.2\lambda$. It is evident that each $d$ value has an optimal set of design parameters that yield the highest realized gain (Table \ref{Table02}).

\begin{table}[!t] 
\centering
\renewcommand{\arraystretch}{1.3} 
\caption{Optimum Design Parameters: Numerical Analysis at 3.5 GHz (Strip Dipoles on Substrate) } 
\label{Table02}
\begin{tabular}{
c
|>{\centering\arraybackslash}m{0.7cm}
|>{\centering\arraybackslash}m{0.7cm}
|>{\centering\arraybackslash}m{0.6cm}
|>{\centering\arraybackslash}m{0.6cm}
|>{\centering\arraybackslash}m{0.6cm}
|>{\centering\arraybackslash}m{0.7cm}
|>{\centering\arraybackslash}m{0.8cm}
}
  \hline
  $d/\lambda$ & $L_1$ (mm) & $L_2$ (mm) & $w_1$ (mm) & $w_2$ (mm) & $\Delta\phi$ (deg.)  & $\eta~(\%)$ & $G_R$ (dBi)\\
  \hline
  \hline
  $0.05$
  &$36.23$
  &$28.89$
  &$2.05$
  &$3.18$
  &$256^{\circ}$
  &$99.2$
  &$3.4$
  \\
  \hline
  $0.1$
  &$42.39$
  &$30.39$
  &$4.85$
  &$3.19$
  &$210^{\circ}$
  &$99.5$
  &$5.0$
  \\
  \hline
  $0.2$
  &$34.27$          
  &$29.48$          
  &$4.40$          
  &$3.19$          
  &$214^{\circ}$                 
  &$99.3$        
  &$6.3$  
  \\
  \hline
  $0.3$    
  &$33.33$          
  &$29.62$         
  &$5.14$          
  &$2.57$          
  &$197^{\circ}$                    
  &$99.5$          
  &$6.0$
  \\
  \hline
  $0.4$    
  &$31.47$          
  &$29.83$          
  &$5.31$          
  &$3.28$          
  &$196^{\circ}$                    
  &$99.6$          
  &$5.2$
  \\
  \hline
  $0.5$    
  &$30.74$          
  &$29.77$         
  &$4.14$          
  &$3.24$          
  &$172^{\circ}$                   
  &$99.7$          
  &$4.1$ 
  \\
  \hline
  \hline
\end{tabular}
\end{table}

When considering the uniform case for $d=0.2\lambda$, the array exhibits a realized gain of $1.1$ dBi and is broadside. This indicates an improvement of approximately $5.2$ dBi for the superdirective array. 

Also, for $d \ge \lambda/2$, the antenna array becomes broadside, and the improvement is marginal, as the optimized antenna realized gain only slightly differs from the uniform case. This aligns with expectations, as the superdirectivity phenomenon does not occur under these conditions (i.e., when $d \ge \lambda/2$).

The radiation efficiency $\eta$ as a function of $d/\lambda$ is also listed in Table \ref{Table02} for inter-element distances up to $0.5\lambda$. It can be observed that the antenna array exhibits extremely high radiation efficiency, exceeding $99.2\%$ for all inter-element cases computed.

Fig. \ref{Fig:RGainVSphi} illustrates the simulated realized gain in the horizontal plane ($H$-plane) for the optimal case with an element spacing of $d=0.2\lambda$. The maximum gain achieved is $6.3$ dBi, observed at $\phi = 0^{\circ}$, indicating an end-fire antenna array configuration. The angular width, estimated at the $3$ dB drop-off points, is $126^{\circ}$. 

\begin{figure}[!t] 
\begin{minipage}{.48\linewidth}
  \centering
  \includegraphics[height=.63\linewidth]{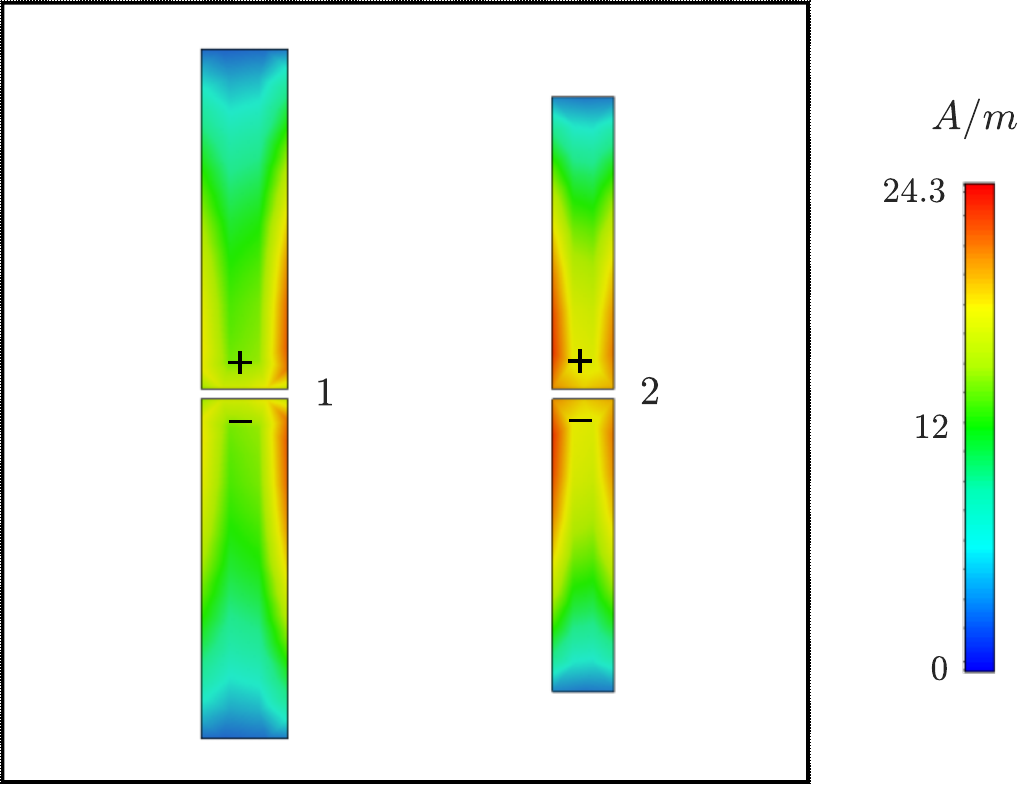}
  \caption{Simulated surface current distribution for simultaneous optimal excitation at $3.5$ GHz, when the inter-element distance is $0.2\lambda$. Ports are numbered, and strip dipoles lie on an RO4003C substrate.}
  \label{Fig:SurfCurrent}
\end{minipage}
\hspace{5mm}
\begin{minipage}{.48\linewidth}
  \centering
  \includegraphics[height=.63\linewidth]{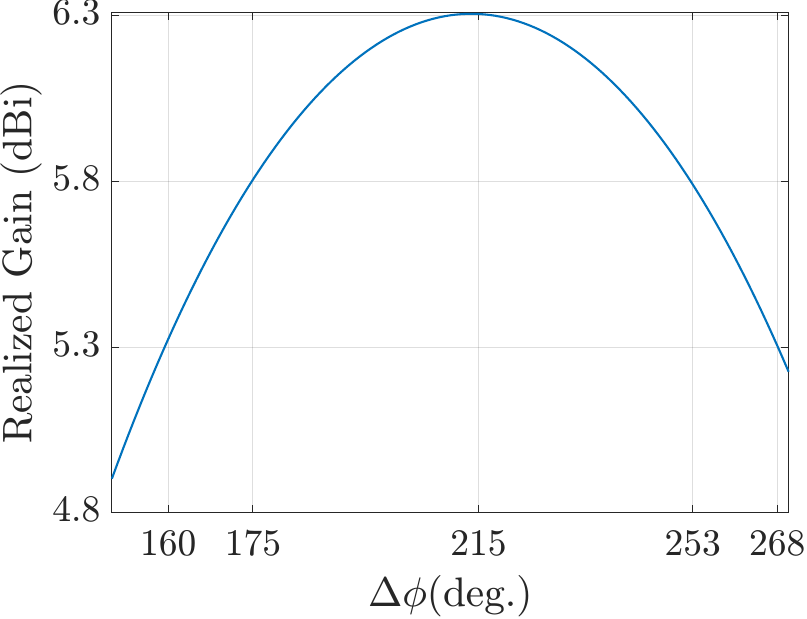}
  \caption{The effect of the phase difference $\Delta\phi$ on the realized gain (simulated results): when $\Delta\phi$ ranges from $175^{\circ}$ to $253^{\circ}$, the realized gain experiences a decrease of merely $0.5$ dB.}
  \label{Fig:MaxRGainVSDeltaPhi}
\end{minipage} 
\end{figure}

In superdirective antenna arrays, the surface current distribution plays a crucial role in achieving high directivity and gain \cite{balanis2015antenna}, and it is typically non-uniform. It is characterized by strong currents flowing in specific regions of the array elements while minimizing currents in other areas. This non-uniform current distribution helps in shaping the radiation pattern and achieving high directivity. The specific current distribution pattern depends on the design and geometry of the array elements. The spacing, size, and arrangement of the elements, as well as the excitation amplitudes and phases, all contribute to the desired surface current distribution. In this work, the simulated surface current distribution is depicted in Fig. \ref{Fig:SurfCurrent}. It is evident that distribution is not uniform, as expected.

The impact of the phase difference $\Delta\phi$ on the realized gain, with all other design parameters at their optimal values, is depicted in Fig. \ref{Fig:MaxRGainVSDeltaPhi}. The graph demonstrates that varying the phase difference to $175^{\circ}$ or $253^{\circ}$ from the maximum at $215^{\circ}$ leads to a reduction in realized gain of $0.5$ dB. This observation is of significance in fabrication, as minor deviations in the phase difference do not exert a substantial influence on the maximum realized gain.

\subsection{Implementation and Measurements}

After completing the theoretical study and conducting a comprehensive numerical analysis using full-electromagnetic simulation, we proceed to the implementation phase. The antenna array is fabricated based on the design concept discussed earlier, which involves strip dipoles with slightly different lengths and radii (or widths). These dipoles are excited by signals of equal magnitude but different phases, enabling impedance matching to $50$ $\Omega$. 

For validation purposes, the antenna array with an inter-element distance of $d = 0.2\lambda$ was selected. To achieve a precise phase difference, coaxial cables (RG405) of different lengths were used to feed the strip dipoles. The strips were etched onto an RO4003C substrate with a thickness of $0.813$ mm, a dielectric constant of $\epsilon_r=3.55$, and a dissipation factor of $\tan\delta=0.0027$. The substrate had dimensions of $50 \times 50$ mm.
A balun (balanced-to-unbalanced) of length $\lambda/4$ was incorporated to match the balanced structure of the dipole to the unbalanced structure of the coaxial cable. SMA connectors were employed for the feeding.
Before the fabrication and measurement, the antenna array was subjected to numerical simulation to determine the optimal design parameters. Similar to the previous design, the lengths $L_1$, $L_2$, $w_1$, and $w_2$ were considered. However, this time, to achieve the optimal phase difference, the length of the first coaxial cable ($L_{c1}$) was fixed at $L_{c1}=\lambda/2$, and the optimal length of the second cable ($L_{c2}$) was determined. Following numerical optimization, the final design parameters were obtained: $L_1=33.7$ mm, $L_2=29.1$ mm, $w_1=4.8$ mm, $w_2=3.5$ mm, and $L_{c2}=81.3$ mm.

\begin{figure}[!t] 
\centering
  \includegraphics[height=0.2\linewidth]{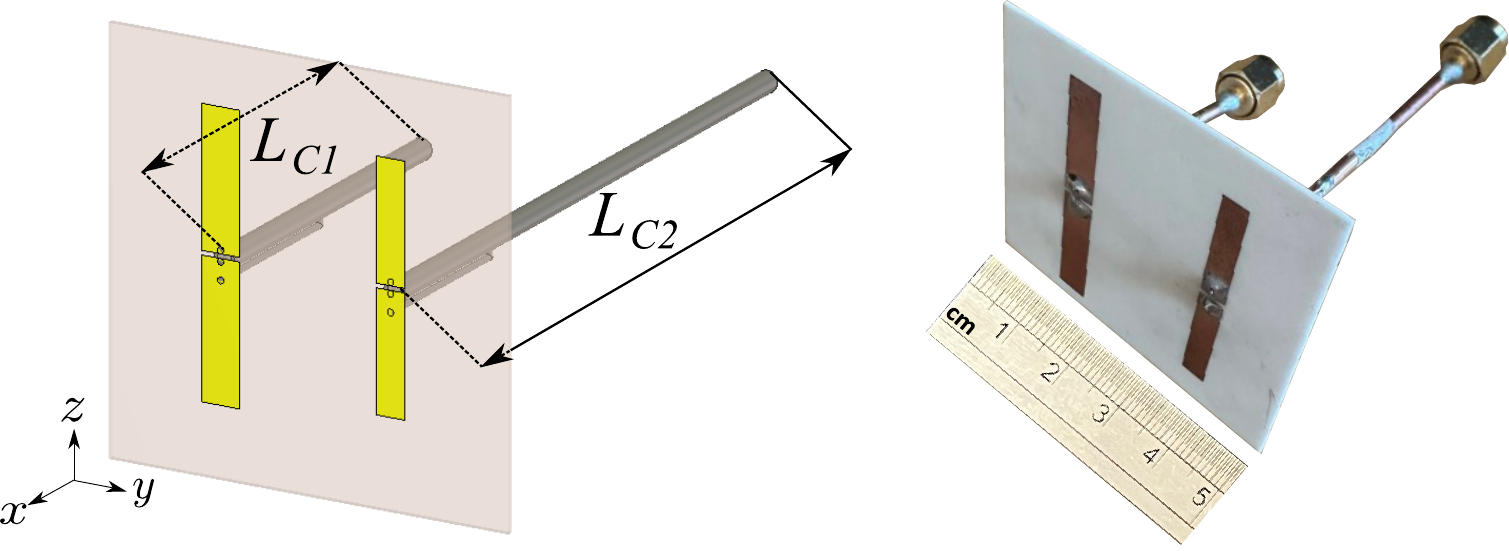}
  \caption{(left) Prior to fabrication, the antenna array underwent simulation to obtain optimal values for the design parameters. In this scenario, the inter-element phase difference was achieved by utilizing coaxial cables of different lengths: $L_{c1}$ was fixed at $\lambda/2$, while $L_{c2}$ was estimated as a design parameter instead of $\Delta\phi$. Additionally, a pawsey balun with a length of $\lambda/4$ was implemented on both dipoles to ensure a smooth transition from the coaxial cables to the strip dipoles. (right) Fabricated antenna array.}
  \label{Fig:FabGeom}
\end{figure}

\begin{figure}
\begin{minipage}{.48\linewidth}
  \centering
  \includegraphics[height=.63\linewidth]{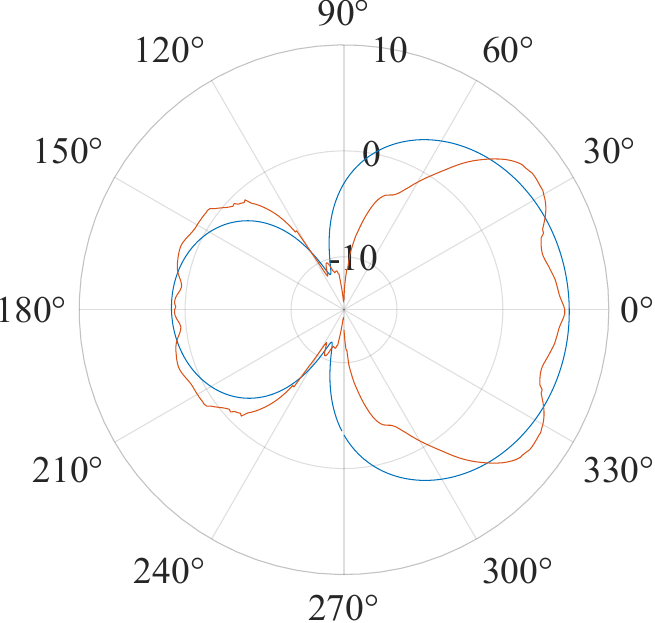}
  \caption{Measured and simulated results of the realized gain in the horizontal plane ($H$-plane): indicating that the proposed antenna is behaving as a super realized gain antenna array, with a maximum realized gain of $6.3$ dBi, at $3.5$ GHz.}
  \label{Fig:RGainSimVSMsr}
\end{minipage} 
\hspace{5mm}
\begin{minipage}{.48\linewidth}
  \centering
  \includegraphics[height=.60\linewidth]{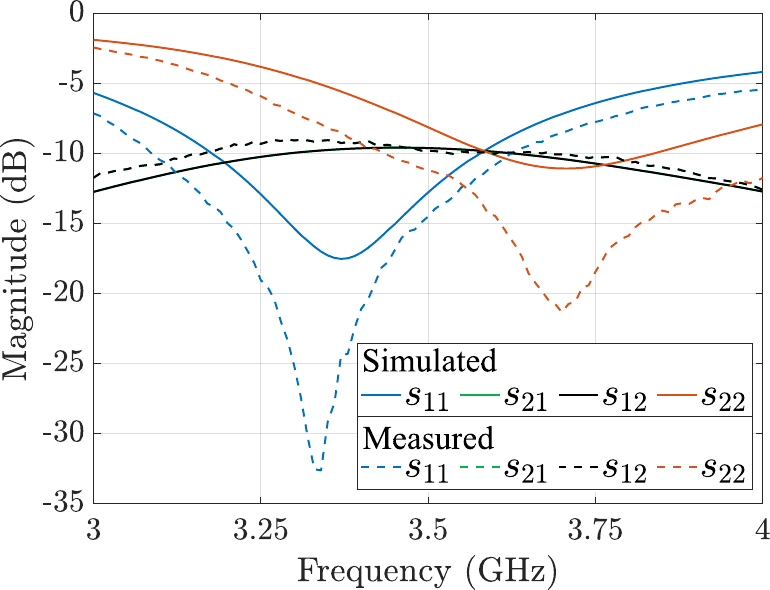}
  \caption{Measured and simulated S-parameters of the fabricated antenna array. It is evident that the reflection coefficient at both dipoles is less than $-10$ dB, indicating a well impedance-matched to $50$ $\Omega$.}
  \label{Fig:Sparam}
\end{minipage}
\end{figure}

To measure the realized gain, we employed the method of three antennas \cite{balanis2015antenna}. This method involves the use of a transmitting antenna, the antenna under test, and a reference antenna. In our setup, we opted to use identical transmitting and reference antennas \cite{drh118} to simplify calculations. A signal generator was utilized at the transmitter, emitting at a frequency of $3.5$ GHz with a power of $0$ dBm, while a spectrum analyzer was employed at the receiver. Additionally, we took into account any losses incurred by the cables used in our setup.
Instead of using a power divider, we performed our measurement in two steps and then combined the results algebraically. Specifically, in the first step, we measured the received power at the first dipole while terminating the second dipole at $50$ $\Omega$. In the second step, we repeated the process but with the roles of the first and second dipoles reversed. The antenna under test was rotated to obtain the received power in the horizontal plane ($H$-plane). Due to the antenna's symmetry, we measured the rotation angles from $0^{\circ}$ to $180^{\circ}$.

The measured results, along with the simulated results, are shown in Fig. \ref{Fig:RGainSimVSMsr}. Good agreement between the measured and simulated results is observed. At 3.5 GHz, the proposed antenna exhibits a measured realized gain of $6.3$ dBi. The observed ripples could be due to balun manufacturing tolerances, the finite size of the substrate, or attenuation.

Fig. \ref{Fig:Sparam} depicts the measured and simulated $S$-parameters of the fabricated antenna array, demonstrating a high level of agreement between them. Additionally, at $3.5$ GHz, both $S_{11}$ and $S_{22}$ exhibit values below $-10$ dB, indicating low return losses at $50$ $\Omega$. By analyzing the $S$-parameters, we can estimate the impedance of each antenna element \cite{balanis2015antenna}. Specifically, at $3.5$ GHz, the first and second dipoles have impedance values of $53.5 + j1.2$ and $57.9 + j2$, respectively, the measured system resulting in reflection coefficients of $-14.5$ dB and $-11.2$ dB for dipole 1 and dipole 2, respectively. Based on Fig. \ref{Fig:Sparam}, the antenna array operates (i.e., the reflection coefficient is below $-10$ dB for both dipoles at a $50$ $\Omega$ impedance), within the frequency range of $3.44$ GHz to $3.62$ GHz, resulting in a measured fractional bandwidth of $5.1 \%$.

The measurements demonstrate a close alignment with the simulated and theoretical results, confirming the successful realization of super realized gain. This validates the efficacy of our approach in reducing losses, increasing radiation efficiency, and enhancing the power efficiency of the superdirective antenna system.

\balance

\section{Conclusion}

This study undertook a thorough investigation of a superdirective dipole antenna array specifically designed to cater to the requirements of directional 5G wireless communication applications. The research focused on careful design considerations, including the selection of appropriate radiating elements, to achieve the desired superdirective performance. Impedance matching was effectively addressed through adjustments in strip dimensions, leveraging innovative techniques not previously introduced in the literature. Phase matching was ensured by employing unequal coaxial cable lengths to achieve the desired inter-element phase difference. Furthermore, the antenna array's efficiency was optimized by minimizing both ohmic and return losses through appropriate low-profile antenna design, resulting in high radiation efficiency. The measured and simulated results exhibited exceptional agreement, confirming the effectiveness of the proposed design. This work contributes to the advancement of high-directivity antennas and provides valuable insights for future research and development in the field of superdirective antenna arrays.

\bibliographystyle{IEEEtran}
\bibliography{mybib.bib}

\balance

\end{document}